\documentclass[10pt,twocolumn,preprintnumbers,superscriptaddress,nofootinbib,aps,prd]
{revtex4-2}

\usepackage{graphicx}
\usepackage{amsmath}
\usepackage{srcltx}
\usepackage[utf8]{inputenc}
\usepackage[colorlinks]{hyperref}
\usepackage{url}
\usepackage{times}
\usepackage{amssymb}
\newcommand{\vev}[1]{\langle {#1} \rangle}
\newcommand{\lsim}{\lesssim}
\newcommand{\gsim}{\gtrsim}

\newcommand{\eq}[1]{Eq.~(\ref{#1})}

\newcommand{\ord}[1]{\mathcal{O}{(#1)}}
\newcommand{\beq}{\begin{equation}}
\newcommand{\eeq}{\end{equation}}
\newcommand{\bea}{\begin{eqnarray}}
\newcommand{\eea}{\end{eqnarray}}

\newcommand{\appropto}{\mathrel{\vcenter{
  \offinterlineskip\halign{\hfil$##$\cr
    \propto\cr\noalign{\kern2pt}\sim\cr\noalign{\kern-2pt}}}}}

\begin{document}

\pagestyle{plain}

\title{\boldmath Bringing Peccei-Quinn Mechanism Down to Earth}

\author{Hooman Davoudiasl}
\email{hooman@bnl.gov} 
\author{Marvin Schnubel}
\email{mschnubel@bnl.gov}

\affiliation{High Energy Theory Group, Physics Department \\ Brookhaven National Laboratory,
Upton, NY 11973, USA}


\begin{abstract}

It is conventionally assumed that the physics underlying the Peccei-Quinn (PQ) mechanism for addressing the strong CP problem is at very high energies, orders of magnitude above the weak scale.  However, this may not be the case in general and the associated PQ boson  $\phi$, besides the signature state, {\it i.e.} the ultralight axion $a$, may emerge well below the weak scale.  We consider this possibility and examine some of the conditions for its viability.  The example model proposed here may also provide the requisite Standard Model Higgs mass parameter, without invoking new scalars above the GeV scale.  The corresponding parameter space can maintain {\it finite naturalness} against quantum corrections. This scenario, depending on choice of parameters, can potentially be constrained by flavor data.  We point out that the current mild excess in $B^+\to K^+ \nu \bar \nu$, reported by the Belle II experiment, could be explained in this setup as $B^+\to K^+ \phi$ and $B^+\to K^+ a$, with both $\phi$ and $a$  escaping the detector as missing energy.  For a sufficiently heavy PQ boson, in the GeV regime,  one can separate these two contributions, due to the difference in $K^+$ momenta.  In this case, the axion may also affect lighter meson, {\it e.g.} kaon, decays while $\phi$ would not be a kinematically allowed final state.

\end{abstract}
\maketitle


One of the lingering mysteries of modern particle physics is that the strong interactions of hadrons seem to respect the combined charge conjugation and parity symmetry (CP) at a high level of precision.  This is especially intriguing given that quantum chromodynamics (QCD) -- the microscopic theory of strong interactions -- can in principle allow for CP violation at the renormalizeable level, parameterized by a constant $\bar\theta$ that may {\it a priori} be $\ord{1}$.  However, current experimental bounds on neutron electric dipole moment, a CP violating observable, imply that $\bar\theta\lsim 10^{-10}$ \cite{Abel:2020pzs}.  Since weak interactions are known to violate CP, there is no ground for assuming that it is a good fundamental symmetry, in order to justify setting $\bar\theta = 0$.  The puzzling smallness of this parameter of QCD is often referred to as the {\it strong CP problem}. 

Of the proposed resolutions of the above puzzle, perhaps the most intensely investigated one, by both theory and experiment, is the Peccei-Quinn (PQ) mechanism \cite{Peccei:1977hh,Peccei:1977ur}.  The PQ idea is to turn $\bar\theta$ into a dynamical field that starts out being a massless Goldstone boson, emerging from the spontaneous breaking of a global $U(1)_{\rm PQ}$ symmetry \cite{Weinberg:1977ma,Wilczek:1977pj}.  This symmetry has an anomaly under the $SU(3)_c$ gauge group of strong interactions, which constitutes an explicit breaking of $U(1)_{\rm PQ}$.  Therefore, non-perturbative QCD dynamics can generate a potential for the aforementioned Goldstone boson, referred to as the ``axion" $a$, giving it a small non-zero mass $m_a$ \cite{Shifman:1979if,Bardeen:1978nq,DiVecchia:1980yfw}. 

The strength of the axion interactions with the Standard Model (SM) particles is inversely proportional to its decay constant $f_a$, which is set by the vacuum expectation value (vev) of the scalar $\Phi$ that spontaneously breaks $U(1)_{\rm PQ}$, {\it i.e.} $\vev{\Phi} = f_a/\sqrt{2}$ \cite{Ringwald:2012hr}.  Due to the stringent limits on the strength of the axion interactions with various particles, $f_a$ has been pushed to very high scales: $f_a\gsim 10^8$~GeV \cite{Davidson:1981zd,Davidson:1984ik,Peccei:1986pn,Krauss:1986wx,Geng:1988nc,Celis:2014iua,Alves:2017avw,DiLuzio:2017ogq,Choi:2017gpf,MartinCamalich:2020dfe,Gelmini:1982zz,Anselm:1985bp,Bauer:2017nlg,Bauer:2017ris,Bauer:2021wjo,ParticleDataGroup:2022pth}.

Given the above account, it is generally assumed that PQ symmetry breaking is described by {\it ultraviolet} physics at energies of $\ord{f_a}$, which is inaccessible to low energy experiments.  In particular, the mass parameter of $\Phi$ is generically expected to be quite large; $m_\Phi\gsim 10^8$~GeV.  In this work, we will challenge this expectation and examine conditions under which the physics that generates the QCD axion decay constant may give rise to particles at mass scales of $\ord{\rm GeV}$. Recently, a different study showed that light scalars can also arise in Nelson-Barr type solutions to the strong CP problem \cite{Dine:2024bxv}. Various conceptual ingredients employed in our discussion can be found in earlier works.  Yet, we will try to place the general ideas in a different context, with new implications which can potentially be accessible to measurements.  We will next describe a simple setup that can demonstrate the above picture.  

Consider the scalar potential
\bea\nonumber 
V(H, \Phi) &=& \frac{\lambda_H}{2} (H^\dagger H)^2 - \kappa\, (H^\dagger H)(\Phi^\dagger \Phi)\\ 
&+& \frac{\lambda_\Phi}{2} (\Phi^\dagger \Phi)^2 - 
\mu_\Phi^2\, \Phi^\dagger \Phi\,,
\label{VHPhi}
\eea
where $\lambda_H$, $\lambda_\Phi$, and $\kappa$ are constants and the mass parameter $\mu_\Phi^2 >0$.  One can rewrite the above potential in the form  
\bea\nonumber
V(H, \Phi) &=& \frac{\lambda_H}{2} 
\left[|H|^2 - \sqrt{\frac{\lambda_\Phi}{\lambda_H}}\,|\Phi|^2\right]^2\\ 
&+& \left(\sqrt{\lambda_H \lambda_\Phi} - \kappa\right)|H|^2 |\Phi|^2 - 
\mu_\Phi^2 |\Phi|^2\,,
\label{VHPhiII}
\eea
where $|S|^2 \equiv S^\dagger S$ for the scalar $S=H,\Phi$. From \eq{VHPhiII}, we can infer that the classical stability of the potential $V(H,\Phi)$ against runaway behavior requires 
\beq
\lambda_H > 0\quad ; \quad \lambda_H \lambda_\Phi > \kappa^2.
\label{quartics}
\eeq
Henceforth, we will assume that $\lambda_H, \lambda_\Phi, \kappa >0$.  

In order to generate a potential for the axion via QCD dynamics, it needs to couple to the gluon field.  We can arrange for this through the Yukawa coupling to $\Phi$ of a new color charged fermion $F$. For concreteness, we will assume $F$ to have the quantum numbers of an $SU(2)$ singlet down quark in the SM, but with chiral charge $Q$ assignment under the $U(1)_{\rm PQ}$ symmetry: $Q(F_L) = Q(\Phi) = +1$ and $Q(F_R)=0$.  We can then write down     
\beq
y_F\, \Phi \bar F_L F_R + \text{\small H.C.},
\label{Yukawa}
\eeq
where $y_F \in\mathbb{R}$.  We may decompose the scalar $\Phi$ into its radial $\phi$ and angular $a$ parts 
\beq
\Phi = \left(\frac{f_a + \phi}{\sqrt{2}}\right) e^{i a/f_a}.
\label{Phi}
\eeq
From \eq{Yukawa}, we see that the $U(1)_{\rm PQ}$ symmetry breaking through $\Phi$ gives $F$ a mass 
\beq
m_F=\frac{y_F}{\sqrt{2}} f_a.    
\label{mF}
\eeq
We also recover the usual axion interaction 
\beq
i g_a a \bar F\gamma_5 F\,,
\label{ga}
\eeq 
where $g_a = m_F/f_a$.  This interaction will lead to a coupling between the axion and the gluon  
\beq
\sim \left(\frac{\alpha_s}{8\pi}\right) \frac{a}{f_a}\,G_{\mu\nu}{\tilde G^{\mu\nu}}\,,
\label{axion-gluon}
\eeq 
where $\alpha_s$ is the strong interaction fine structure constant, $G_{\mu\nu}$ is the gluon field strength tensor, and $\tilde{G}^{\mu\nu}$ is its dual.  

Given the above, one would obtain the QCD contribution to the axion potential, and the PQ dynamical resolution of the strong CP problem.  Since all of that physics is very well-studied in a number of references, we will not review it here further.  Instead, we will examine what conditions are implied by the above setup if  $\Phi$ is a light boson.  What follows is not a comprehensive study, but we aim to include sufficient details to establish the feasibility of the framework.  The numerical treatment will be at the order-of-magnitude level, since we do not advocate for a specific realization of the model with precisely set parameters.  Nonetheless, we will adopt a set of reference model parameters, up to implied $\ord{1}$ factors, to demonstrate our main points.  The interested reader would be able to extend our analysis, straightforwardly, to other regions of parameter space if desirable.

Henceforth, let us take $f_a=10^{9}$~GeV, consistent with all existing bounds on ultralight axions.  As $F$ has color charge, we must ensure that its mass is sufficiently large so that it is not ruled out by current LHC data.  A precise bound depends on the couplings of $F$ to other fields.  However, we will take $m_F\gsim 3$~TeV to be large enough and safe \cite{ParticleDataGroup:2022pth}, corresponding to $y_F \gsim 4 \times 10^{-6}$, due to lack of a signal in direct searches.  Yet, such mass scales can still be constrained by precision flavor physics data, as we will discuss later.

There is no obvious choice for the mass of $\phi$, the radial component of $\Phi$.  However, let us set $m_\phi =  1$~GeV, well below the weak scale, which would  reasonably designate  $\phi$ a low energy degree of freedom, as intended in this work.  From \eq{VHPhi}, we can infer 
\beq
m_\phi^2 = \lambda_\Phi f_a^2\,,
\label{mrho2}
\eeq
which implies $\lambda_\Phi =  10^{-18}$, for our choice of $m_\phi$.     

The observed Higgs mass $m_H \approx 125$~GeV \cite{ParticleDataGroup:2022pth} yields a tachyonic mass parameter $\mu_H = m_H/\sqrt{2}\approx 90$~GeV.  With our choice of scalar potential in \eq{VHPhi}, we see that $\mu_H^2 = \kappa f_a^2/2$; we have implicitly assumed that a ``bare" dimension-2 Higgs mass term is absent or negligible compared to this contribution.  We thus obtain 
\beq
\kappa \approx 2 \times 10^{-14} \left(\frac{10^9~{\rm GeV}}{f_a}\right)^2.
\label{kappa}
\eeq
In this picture, electroweak symmetry breaking can be induced after PQ symmetry breaking, in the early Universe.  Having fixed all the parameters of the scalar potential used in our scenario, we next examine the consistency and stability of the obtained values. We note that for our benchmark parameters, to a good approximation $\mu_\Phi = m_\phi/\sqrt{2} \approx 0.7$~GeV.  The mixed quartic can also contribute to $\mu_\Phi^2$, once the Higgs gets a vev.  However, for $\vev{H}=v_H/\sqrt{2}$, with $v_H\approx 246$~GeV, that contribution is $\lsim (\text{25 keV})^2$ and hence negligible compared to the above value of $\mu_\Phi^2$.

Given that $\lambda_H$ is $\ord{10^{-1}}$, we have $\lambda_H \gg \kappa$ and hence the second term in \eq{VHPhi} cannot have any significant quantum contribution to it.  However, the interaction proportional to $\kappa$ can generate a 1-loop contribution to the $\Phi$ quartic coupling
\beq
\delta \lambda_\Phi(\kappa) \sim \frac{\kappa^2}{16 \pi^2} \ln\left(\frac{v_H^2}{\mu_\Phi^2}\right)\,.
\label{del-lamPhi-kappa}
\eeq
Here and in what follows we will only consider finite loop contributions set by physical parameters, ignoring divergent parts of the loop which are removed by a regulator.   For our reference values, this yields $\delta \lambda_\Phi\lsim 10^{-29}\ll \lambda_\Phi$.

Next, let us consider the contributions from the Yukawa interaction in \eq{Yukawa}.  This term can result in a 1-loop contribution to the $\Phi$ quartic given by
\beq
\delta \lambda_\Phi(y_F) \sim \frac{y_F^4}{16 \pi^2} \ln\left(\frac{m_F^2}{\mu_\Phi^2}\right)\,.
\label{del-lamPhi-yF}
\eeq 
To maintain quantum stability of our parameter space, we require $\delta \lambda_\Phi(y_F) \lsim \lambda_\Phi$.  This yields $y_F\lsim 5\times 10^{-5}$.  We have used $m_F\sim \ord{{\rm 10~TeV}}$, as a typical choice for our study.  The corresponding quantum correction to $\mu_\Phi^2$ can be estimated by 
$\delta \mu_\Phi^2 \sim y_F^2 m_F^2/(16\pi^2)\ll(\text{0.7 GeV})^2$, which is consistent with our reference value $\mu_\Phi^2 \approx (\text{0.7 GeV})^2$.  For our reference parameter, we will consider $y_F = 10^{-5}$, in light of the above discussion.

The preceding  analysis shows that, so far, our choice of parameters are consistent and stable against quantum corrections.  In particular, the Higgs mass does not get any large contributions proportional to new physical scales and hence it satisfies {\it finite naturalness} \cite{Farina:2013mla}.  To see this, we observe that any such loop contribution in our model would be at most $\kappa \mu_\Phi^2/(16\pi^2) \lsim 10^{-16}$~GeV$^2$ and hence completely negligible.  We have collected the reference parameter values of our scenario in Table~\ref{t1}.   
\vskip0.2cm
\begin{table}
\begin{tabular}{|c|c|c|c|c|c|}
	\hline
	parameter	&  $\kappa$ & $\lambda_\Phi$ & $\mu_\Phi$  & $f_a$ & $y_F$ \\
	\hline
	value	& $2\times 10^{-14}$ & $10^{-18}$  & 0.7~GeV & $10^{9}$~GeV & $10^{-5}$ \\
	\hline
\end{tabular}  
\caption{Reference values of model parameters used in this work.}
\label{t1}
\end{table}

Having set the parameters of the model, let us now consider its phenomenology.  The QCD axion physics would follow and its phenomenology is the same as in the conventional scenarios\footnote{We refer the interested reader to Ref.~\cite{Ringwald:2024uds} for a recent review}. The fermion $F$ can be potentially accessible to the LHC (for somewhat smaller masses than taken as our benchmark), or else a future high energy hadron collider, such as FCC-hh with a center of mass energy of 100~TeV \cite{FCC:2018vvp}. 

The most novel aspect of our scenario is the low mass of the remnant PQ breaking boson, $\phi$.  One could roughly estimate the coupling of $\phi$ to gluons, induced by it interactions with $F$, as \cite{Knapen:2017xzo} 
\beq
\frac{\alpha_s}{2\sqrt{2}\,\pi\,f_a} \phi\,G^a_{\mu\nu}G^{a\mu\nu}\,,
\label{rhoGG}
\eeq
where $\alpha_s$ is the QCD coupling constant and $G^a_{\mu\nu}$ is the gluon field strength tensor; $a=1,2,\ldots,8$.  This yields a coupling to nucleons
\beq
y_N \, \phi \bar N N\,,
\label{rhonn}
\eeq
where \cite{Knapen:2017xzo} 
\beq
y_N \approx 2.7\times 10^{-10}\left(\frac{10^{9}\,\text{GeV}}{f_a}\right).
\label{yN}
\eeq
The decay of $\phi$ into gluon pairs, for $m_\phi \gsim 1$~GeV, is given by 
\beq
\Gamma(\phi\to gg) = \frac{\alpha_s^2 \, m_\phi^3}{4\,\pi^3 f_a^2}\,.
\label{Gam-gg}
\eeq

The coupling of $\phi$ to photons, mediated by $F$ at 1-loop, is given by 
\beq 
\frac{g_\gamma}{4} \phi F_{\mu\nu}F^{\mu\nu}\,,
\label{rhogamgam}
\eeq
with (see, {\it e.g.}, Ref.~\cite{Davoudiasl:2018fbb})
\beq
g_\gamma = \frac{2 \sqrt{2}\, \alpha\, Q_F^2\, N_c^F}{3 \pi\, f_a}\,,
\label{ggam}
\eeq
where $Q_F$ is the electric charge of $F$ and $N_c^F$ is its number of colors.  For our reference values, we get $g_\gamma\sim  7 \times 10^{-13}$~GeV$^{-1}$.  The partial width for $\phi\to \gamma \gamma$ is then given by
\beq
\Gamma(\phi\to \gamma \gamma) = \frac{g_\gamma^2\, m_\phi^3}{64\,\pi}\,.
\label{Gam-2photon}
\eeq
Using the reference values as before, we find $\Gamma(\phi\to gg) \approx 2 \times 10^{-21}$~GeV and $\Gamma(\phi\to \gamma \gamma) \sim 2 \times 10^{-27}$~GeV, where we have used $\alpha_s(m_\phi)\approx 0.5$, for $m_\phi=1$~GeV. The decay rate of $\phi$ into light hadrons can be reasonably well approximated by its decay rate into a pair of gluons. Hence the lifetime of $\phi$ is roughly estimated  by $\sim 3 \times 10^{-4}$~s.  The decay length of $\phi$ is then $\sim 10^5$~m for our reference values, making it a missing energy signal in high energy experiments.  

Here, we would like to mention that the $\Phi$ coupling to Higgs in \eq{VHPhi} will lead to the mixing of the Higgs boson $h$ with $\phi$.  This mixing is given by the angle
\beq
\theta_{\phi h}\approx \frac{\kappa v_H f_a}{4 m_h^2}\approx 7.8 \times 10^{-8} \left(\frac{f_a}{10^9~\text{GeV}}\right),
\label{phih-mix}
\eeq
where we have used the reference value of $\kappa$ and $m_h\approx 125.2$~GeV.  Hence, $\phi$ would couple to all SM fermions through mixing with the Higgs, suppressed by $\theta_{\phi h}$.  This implies a width for $\phi\to \mu^+\mu^-$ given by
\beq
\Gamma (\phi \to  \mu^+\mu^-) \approx \frac{\theta_{\phi h}^2\, m_\mu^2 m_\phi}{8\pi v_H^2}\,,
\eeq
where $m_\mu \approx 0.106$~GeV is the muon mass.  With our benchmark parameters we find $\Gamma (\phi \to  \mu^+\mu^-) \approx 4.6 \times 10^{-23}$~GeV and hence the lifetime of $\phi$ for $m_\phi$ near the GeV scale is set by its decay into gluons, discussed earlier.  

We also add that in the model described by \eq{VHPhi}, there are additional contributions to the coupling of $\phi$ to gluons, mediated by its mixing with the Higgs, through heavy quark loop diagrams.  However, for a GeV scale $\phi$, one can show that these contributions are $\sim 3/4$ of the value induced by the coupling to $F$.  To see this, note that the coupling to gluons through Higgs mixing $\propto 3\theta_{\phi h}/v_H$, summing over top,  bottom, and charm quarks, while the one generated by coupling to $F$ is proportional to $1/f_a$.  Using \eq{phih-mix} one can then show that the ratio of these couplings is $\sim 3 \kappa f_a^2/(4 m_h^2)$, which is $\sim 3/4$.  Hence, this contribution does not change the qualitative features of our scenario.

\begin{figure}
\includegraphics[width=0.2\textwidth]{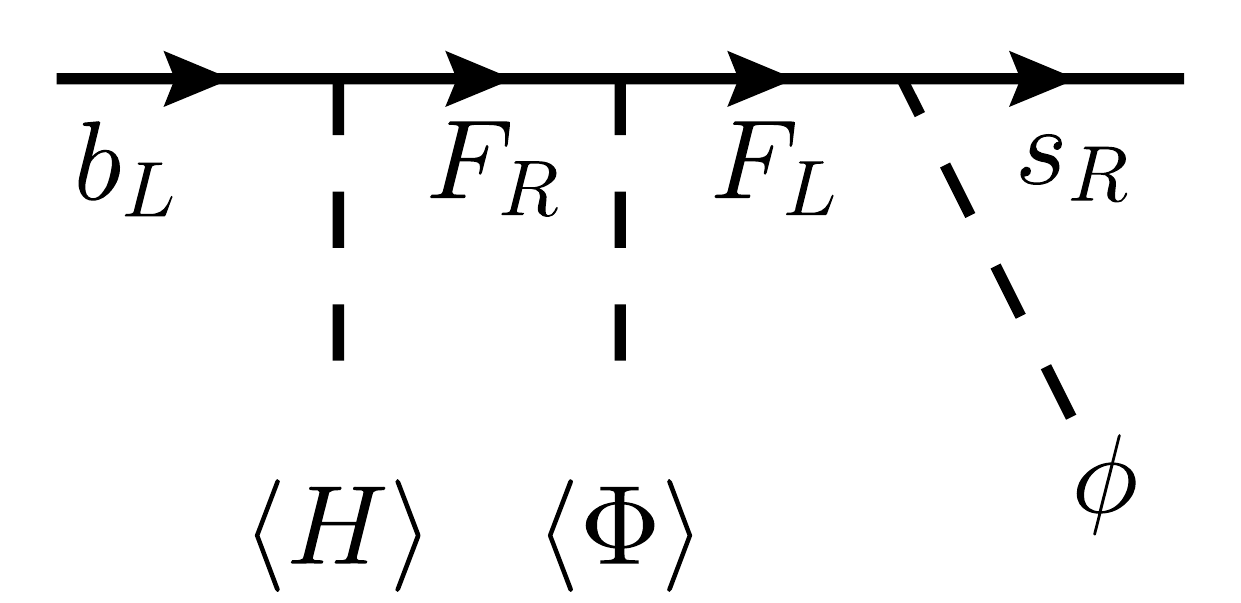}
\centering
\caption{Feynman diagram showing how an effective $b\to s\phi$ transition is mediated by mixing of SM quarks with the heavy PQ fermion $F$. Here, $\langle H\rangle$ and $\langle\Phi\rangle$ denote insertions of the Higgs and PQ scalar vev, respectively.}\label{fig:effbs}
\end{figure}

Assuming $F$ has the quantum numbers of a down-type quark denoted by $q_R$, the following terms are generally allowed in the Lagrangian
\begin{equation}\label{eq:Lagterms}
    \mathcal{L}\supset y_F\Phi \bar{F}_LF_R+y_{Fq}\Phi\bar{F}_L q_R+y_{QF} H \bar{Q}_LF_R+\text{\small H.C.}\,,
\end{equation}
where generation indices have been suppressed. Effective flavor-changing neutral transitions like $b\to s\phi$ and $b \to s a$, for $Q_L = (t_L, b_L)$ the third generation quark doublet and $q_R= s_R$ the right-handed strange quark, are consequently generated through diagrams like the one shown in figure~\ref{fig:effbs}. 
The corresponding effective Lagrangian reads   
\begin{equation}
\mathcal{L}^\text{eff}\supset \frac{C_{bs}}{2\sqrt{2}}(\phi + i a)\,\bar{b} (1 + \gamma_5) s +\text{\small H.C.}\,,
\end{equation}
with 
\beq
C_{Qq} \approx \frac{y_{Fq}y_{QF}\vev{H}}{m_F}\,,
\label{CQq}
\eeq
which yields 
$C_{bs} \approx 3\times10^{-2}(y_{Fs}y_{QF})_{bs}$. Constraints on this and similar coupling  combinations come from searches for flavor-violating $B$ and $K$ meson decays.  Note that due to the chiral nature of the couplings in \eq{eq:Lagterms}, $\phi$ and $a$ have both scalar and pseudo-scalar flavor-violating couplings. 

By symmetry, the $B\to K$ matrix element for the $\bar b \gamma_5 s$ interaction vanishes \cite{Bolton:2024egx}. 
The decay rate for $B\to K \varphi$, with $\varphi=a,\phi$, is given by \cite{Knapen:2017xzo,Willey:1982mc,Bolton:2024egx} 
\begin{equation}
\begin{aligned}
    \Gamma(B\to K\varphi)=&\frac{|C_{bs}|^2}{128\pi m_B^3}\frac{(m_B^2-m_K^2)^2}{(m_b-m_s)^2}\\
    &\times\lambda^{\frac{1}{2}}(m_B^2,m_K^2,m_{\varphi}^2)f_0^2(m_\varphi^2)\,.
    \end{aligned}
    \label{BtoK}
\end{equation}
Here, $\lambda(x,y,z)=(x-y-z)^2-4yz$ is the K\"{a}ll\'{e}n function, $f_0(q^2)=r_2/(1-q^2/m_\text{fit}^2)$ with $r_2=0.33$ and $m_\text{fit}^2=37.46$~GeV$^2$ is a parameterization of the hadronic form factor \cite{Ball:2004ye}.   
In the above, $m_B\approx 5.28$~GeV, $m_K\approx 0.494$~GeV, $m_b\approx 4.18$~GeV and $m_s\approx 0.093$~GeV are the masses of the $B^+$ and $K^+$ mesons and the $b$ and $s$ quarks, respectively.  Below, we will also consider possible constraints from $B\to K^* \varphi$ decays.  Here, only the pseudo-scalar current contributes.  We have \cite{Bolton:2024egx}
\begin{equation}
\begin{aligned}
    \Gamma(B\to K^*\varphi)=&\frac{|C_{bs}|^2}{128\pi m_B^3(m_b-m_s)^2}\\
    &\times\lambda^{\frac{3}{2}}(m_B^2,m_{K^*}^2,m_{\varphi}^2)A_0^2(m_\phi^2)\,,
    \end{aligned}
    \label{BtoK*}
\end{equation}
where $m_{K^*}\approx 0.896$~GeV and  $A_0(q^2)$ is a form factor given by
\begin{equation}
A_0(q^2)=\frac{r_1}{1-q^2/m_R^2}+\frac{r_2}{1-q^2/m_\text{fit}^2}
\end{equation}
with $r_1=1.364$, $r_2=-0.99$, $m_R=5.28$~GeV, and $m_\text{fit}^2=36.78$~GeV$^2$ \cite{Ball:2004rg}.

Given the lifetime estimate of $\phi$ from above, it will escape all detectors and therefore be reconstructed as missing energy in experiments. While the recent Belle II measurement of $B\to K\nu\bar{\nu}$ presents the strongest bounds, it is also exceeding the SM expectation by $2.7\sigma$ \cite{Belle-II:2023esi}. The Belle II  collaboration finds 
\begin{equation}
\text{Br}(B\to K\nu\bar{\nu})=[2.3\pm0.5(\text{stat})^{+0.5}_{-0.4}(\text{syst})]\times10^{-5}.
\label{BtoKnunubar}
\end{equation}
In our model, the light PQ boson could potentially provide an explanation of this excess, together with its axion counterpart.  For almost all masses up to $m_\phi\lesssim m_B-m_K$ the effective coupling $C_{bs}$ is constrained to $C_{bs}< 6\times10^{-8}$ at the $2\sigma$ level, summing over both $a$ and $\phi$, using Eqs.~(\ref{BtoK}) and (\ref{BtoKnunubar}).

Constraints on the same coupling $C_{bs}$ also come from the BaBar measurement \cite{BaBar:2013npw}
\beq
\text{Br}(B\to K^\ast\nu\bar{\nu})=[3.8^{+2.9}_{-2.6}(\text{comb})]\times10^{-5}\,.
\label{BtoK*nunubar}
\eeq
From Eqs.~(\ref{BtoK*}) and (\ref{BtoK*nunubar}), we find the corresponding 2$\sigma$ bound  $C_{bs}< 8\times10^{-8}$\footnote{Even when $F$ does not mix with the SM quarks, {\it i.e.} in cases where the second and third term in the Lagrangian \eqref{eq:Lagterms} are absent, a flavor-violating coupling is generated at the multi-loop level. However, we find
\begin{equation}
    C_{bs} \sim \frac{\alpha_s^2 m_b m_t^2 V_{ts}^\ast V_{tb}}{64\pi^3 v_H^2 f_a}\sim 10^{-15}\,,
\end{equation}
where $\alpha_s\approx 0.12$ is the strong coupling constant; $V_{ts}$ and $V_{tb}$ are CKM matrix elements. This value is well below any experimental constraints.}.

\begin{figure}[t]
\includegraphics[width=0.4\textwidth]{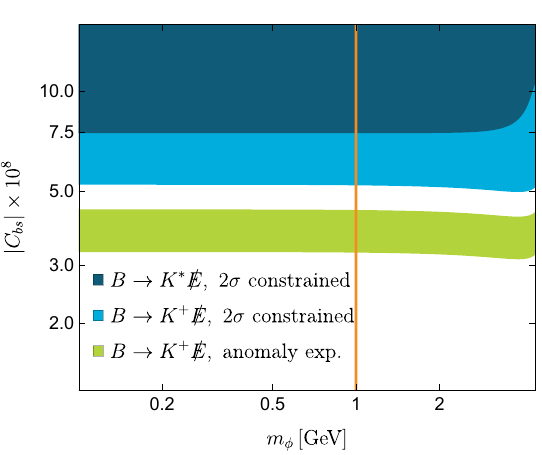}
\centering
\caption{Excluded parameter space of the effective flavor-changing coupling $C_{bs}$ and mass of the light scalar $m_\phi$, from $B\to K^\ast\nu\bar{\nu}$ (dark blue) and $B\to K^+\nu\bar{\nu}$ (light blue), at the $2\sigma$ level. The currently observed excess in the latter channel by Belle-II could potentially be explained within $1\sigma$ of the observed central value (green). We mark our reference value of $m_\phi=1$~GeV with an orange vertical line. The parameter range shown for the mass of the scalar exceeds what we consider a consistent range for our model, which is subject to various parametric conditions as elaborated in the text.}
\label{fig:constraints}
\end{figure}

Choosing couplings $y_{Fs}\lsim 10^{-5}$ and $ y_{QF}\lsim 0.1$, however, evades the above constraints while potentially allowing to explain the observed anomaly. We will assume that the light scalar as well as the axion are solely responsible for the observed difference $\text{Br}(B\to K\nu\bar{\nu})^\text{exp}-\text{Br}(B\to K\nu\bar{\nu})^\text{SM}$ within $1\sigma$. We show the parameter space constrained by the experiment at the $2\sigma$ level as well as the region where the $B\to K\nu\bar{\nu}$ excess is explained by our model within one standard deviation in figure~\ref{fig:constraints} in blue and green, respectively.

Our model is   distinguishable from the SM by a precise determination of the kaon momentum spectrum. While $B\to K\nu\bar{\nu}$ is a three-body decay, $B\to K\varphi$, $\varphi=\phi,a$ as two-body decays would yield a peak in the momentum distribution given at the three-momenta $|\vec{p}_K|=1/(2m_B)\lambda^{(1/2)}(m_B^2,m_K^2,m_\varphi^2)$, and $|\vec{p}_K|^\phi=2.52$~GeV and $|\vec{p}_K|^a=2.62$~GeV, respectively. 

The FCNC transition $b\to d$ with an additional $\phi$ or $a$ radiation is constrained from measurements of $\text{Br}(B\to\pi\nu\bar{\nu})<1.4\times10^{-5}$ (90\% CL) \cite{ParticleDataGroup:2022pth}. This leads to the bound $|C_{bd}|<5\times10^{-8}$ (90\% CL) for our model. This result can be obtained using \eq{BtoK}, by the replacements $K\to \pi$ and $s\to d$, for the corresponding parameters. The fit parameters in the form factor are $r_2=0.258$ and $m_\text{fit}^2=33.81$~GeV$^2$ instead \cite{Ball:2004ye}.

A formula similar to that for $B$ decays can be derived for $K\to\pi a$ decays with the obvious replacements. For our benchmark choice, $\phi$ is too heavy to be produced on-shell in kaon decays. The hadronic form factor in this case is well approximated by $f_0(q^2)\approx 1$ \cite{Marciano:1996wy}, and we have\footnote{Contributions from couplings of the axion to quarks and gluons are negligible in this case.} \cite{Knapen:2017xzo,Willey:1982mc,Bolton:2024egx,Bauer:2021wjo} 
\begin{equation}
\begin{aligned}
    \Gamma(K\to \pi a)=&\frac{|C_{sd}|^2}{128\pi m_K^3}\frac{(m_K^2-m_\pi^2)^2}{(m_s-m_d)^2}\\
    &\times\lambda^{\frac{1}{2}}(m_K^2,m_\pi^2,m_a^2)\,,
    \end{aligned}
\end{equation}
where $C_{sd}$ is given by Eq. \eqref{CQq} with the replacement $(bs)\to(sd)$. 
Here, $Q_L=(c_L, s_L)$ is the second generation quark doublet and $q_R=d_R$ is the right-handed down quark of mass $m_d\approx 5$~MeV. Using $m_\pi\approx 140$~MeV and ignoring the tiny value of $m_a$, the constraint is then $|C_{sd}|\lesssim 6\times 10^{-13}$, coming from the measurement $\text{Br}(K\to\pi\nu\bar{\nu})<1.78 \times10^{-10}$ ($90\%$ C.L.) \cite{NA62:2020fhy} by the NA62 collaboration.

Note that with the above choice of $y_{Fs}$ the loop-generated correction to $\lambda_\Phi$  
\beq
\delta \lambda_\Phi(y_{Fs}) \sim \frac{N_c^F y_{Fs}^4}{16 \pi^2} \ln\left(\frac{m_F^2}{\mu_\Phi^2}\right)
\label{del-lamPhi-yFs}
\eeq
is small and does not destabilize our reference value.  The 1-loop correction to $\kappa$ is estimated by 
\beq
\delta \kappa \sim  
\frac{N_c^F y_{Fs}^2y_{QF}^2}{16 \pi^2} \ln\left(\frac{m_F^2}{\mu_H^2}\right)\,,
\label{del-kappa}
\eeq
which may require a mild $\sim 10 \%$ tuning to yield the reference value for $\kappa$, used in our work.  We also not that if future Belle II data on $B\to K\nu\bar{\nu}$ go back to good agreement with the SM, or if one ignores this modest deviation, one can assume smaller values for $y_{Fs}$ and $y_{QF}$, which could remove the above mild tuning altogether.   The finite 1-loop contribution to the Higgs mass parameter 
\beq
\delta \mu_H^2 \sim \frac{N_c^F y_{QF}^2 m_F^2}
{16 \pi^2}
\label{delmuH2}
\eeq
can also be near the required value and would not necessitate fine-tuned cancellations.

In Ref. \cite{Bolton:2024egx}, the authors  studied how different light new particles could potentially explain the excess as well. They found that under the assumption that the missing energy is attributed to a single radiated and unobserved particle, a scalar of mass $m_\phi=(2.1\pm0.1)$~GeV provides the best fit to the data. Other models including a light scalar that is used to explain the anomaly can be found in \cite{He:2023bnk,He:2024iju,Ho:2024cwk,McKeen:2023uzo,Datta:2023iln,Berezhnoy:2023rxx,Fridell:2023ssf}.

Our light PQ boson $\phi$ also couples to flavor-conserving SM fermion currents, mediated by interactions in \eq{eq:Lagterms}, via mixing with $F$.  However, we find that meson decay experiments like searches for $\Upsilon(1S)\to\gamma\phi$, where the scalar is reconstructed as missing energy, yield the lower bound $f_a\gsim 10^4~\text{GeV}$ \cite{BaBar:2010eww,Kling:2022uzy}, which is far below our assumed value of $f_a\sim10^9$~GeV. For this bound we take $y_{QF}\sim 0.1$ (like before) and the mixing of $F$ and $b$-quarks through $\Phi$ to be small compared to the $F$ Yukawa-coupling, i.e. $y_{Fb}<y_F$.  We emphasize that violating this assumption would result in $F_L$ and $b_R$ instead of $F_L$ and $F_R$ forming the heavy new fermion and thus would require a redefinition of flavors to identify the correct SM $b$ quark.  

Stellar cooling through emission of new particles of horizontal branch stars and red giants only probe masses of $m_\phi\lesssim10$~keV \cite{Raffelt:1988rx,Raffelt:1996wa,Grifols:1986fc,Grifols:1988fv,Hardy:2016kme}, and therefore provide no constraints on the $m_\phi\sim1$~GeV scalar, however they apply to axion emission in our model. In a similar way, axion cooling bounds from supernova 1987A constrain $f_a \gsim 4\times10^{8}$~GeV \cite{Ishizuka:1989ts,Turner:1987by,Burrows:1990pk,ParticleDataGroup:2022pth} just below our assumed value such that a similar exceptional stellar event in the galactic neighbourhood of the Earth in the future could potentially put our model parameters to the test.

\vskip0.5cm
{\tt Digital data for this work can be found as supplementary material associated with the arXiv submission.}

\vskip0.5cm
\begin{acknowledgments}
This work is supported by the US Department of Energy under Grant Contract DE-SC0012704. M.S. gratefully acknowledges support from the Alexander von Humboldt Foundation as a Feodor Lynen Fellow.
\end{acknowledgments}

\bibliography{PQ-IR-refs.bib}

\end{document}